\documentclass[trackchanges, twocolumn]{aastex63}

\shorttitle{MUSE reveals outflows in NGC 7469}
\shortauthors{Robleto-Or\'us et al.}

\usepackage{graphicx}	
\usepackage{amsmath}	
\usepackage{amssymb}	
\usepackage{bm}		
\usepackage{hyperref}
\usepackage{xcolor}
\usepackage{soul} 

\begin{document}

\title{MUSE reveals extended circumnuclear outflows in the Seyfert 1 NGC 7469}

\correspondingauthor{A. C. Robleto-Or\'us}
\email{ac.robletoorus@ugto.mx}

\author[0000-0002-4216-7138]{A. C. Robleto-Or\'us}
\affiliation{Departamento de Astronom\'ia, Universidad de Guanajuato, Apdo. 144, C.P. 36000 Guanajuato, Gto., Mexico}

\author[0000-0002-8009-0637]{J. P. Torres-Papaqui}
\affiliation{Departamento de Astronom\'ia, Universidad de Guanajuato, Apdo. 144, C.P. 36000 Guanajuato, Gto., Mexico}

\author[0000-0001-8825-3624]{A. L. Longinotti}
\affiliation{Instituto Nacional de Astrof\'isica, \'Optica y Electr\'onica, C.P. 72840,  Tonantzintla, Puebla, Mexico}
\affiliation{Instituto de Astronomia, Universidad Nacional Aut\'onoma de M\'exico, C.P. 04510, CDMX, Mexico}

\author[0000-0002-8322-6333]{R. A. Ortega-Minakata}
\affiliation{Instituto de Radioastronom\'ia y Astrof\'isica, Universidad Nacional Aut\'onoma de M\'exico, C.P. 58089, Morelia, Mich., Mexico}

\author[0000-0001-6444-9307]{S. F. S\'anchez}
\affiliation{Instituto de Astronomia, Universidad Nacional Aut\'onoma de M\'exico, C.P. 04510, CDMX, Mexico}

\author[0000-0003-1577-2479]{Y. Ascasibar}
\affiliation{Departamento de F\'isica Te\'orica,Universidad Aut\'onoma de Madrid, C.P. 28049 Madrid, Spain}

\author[0000-0001-9791-4228]{E. Bellocchi}
\affiliation{Centro de Astrobiolog\'ia (CSIC-INTA), ESAC,  Urb. Villafranca del Castillo, E-28691 Villanueva de la Cañada, Madrid, Spain}

\author[0000-0002-1296-6887]{L. Galbany}
\affiliation{Departamento de F\'isica Te\'orica y del Cosmos, Universidad de Granada, C.P. 18071, Granada, Andaluc\'ia, Spain}

\author[0000-0001-6073-9956]{M. Chow-Mart\'inez}
\affiliation{Instituto de Geolog\'ia y Geof\'isica IGG-CIGEO, Universidad Nacional Aut\'onoma de Nicaragua, C.P. 663, Managua, Nicaragua}
\affiliation{Departamento de Astronom\'ia, Universidad de Guanajuato, Apdo. 144, C.P. 36000 Guanajuato, Gto., Mexico}

\author[0000-0003-2293-1802]{J. J. Trejo-Alonso}
\affiliation{Facultad de Ingenier\'ia, Universidad Aut\'onoma de Quer\'etaro, C.P. 76010, Santiago de Quer\'etaro, Qro., Mexico}

\author[0000-0002-9137-861X]{A. Morales-Vargas}
\affiliation{Departamento de Astronom\'ia, Universidad de Guanajuato, Apdo. 144, C.P. 36000 Guanajuato, Gto., Mexico}

\author[0000-0003-0962-8390]{F. J. Romero-Cruz}
\affiliation{Departamento de Astronom\'ia, Universidad de Guanajuato, Apdo. 144, C.P. 36000 Guanajuato, Gto., Mexico}
\affiliation{Instituto Tecnol\'ogico Superior de Guanajuato. C. P. 36262, Guanajuato, Gto. Mexico}

\begin{abstract}
NGC 7469 is a well known Luminous IR Galaxy, with a circumnuclear star formation ring ($\sim 830$ pc radius) surrounding a Seyfert 1 AGN. Nuclear unresolved winds were previously detected in X-rays and UV, as well as an extended biconical outflow in IR coronal lines. We search for extended outflows by measuring the kinematics of the $\mathrm{H\beta}$ and [\ion{O}{3}] $\lambda 5007$ optical emission lines, in data of the VLT/MUSE integral field spectrograph. We find evidence of two outflow kinematic regimes: one slower regime extending across most of the star formation ring---possibly driven by the massive star formation---and a faster regime  (with a maximum velocity of $-715 \ \mathrm{km \ s^{-1}}$), only observed in [\ion{O}{3}], in the western region between the AGN and the massive star forming regions of the ring, likely AGN-driven. This work shows a case where combined AGN/star-formation feedback can be effectively spatially-resolved, opening up a promising path toward a deeper understanding of feedback processes in the central kiloparsec of AGN.
\end{abstract}


\section{Introduction}

Studies over large samples of galaxies have revealed ionized gas outflows associated both with star formation (SF) \citep[e.g.][]{2014_Ho, 2015_Roche, 2017b_Lopez-Coba, 2019_Lopez-Coba} and with active galactic nuclei (AGN) \citep[e.g.][]{2005_Greene, 2016_Woo, 2017_Perna, 2020_Wylezalek}, significantly improving our understanding of the role of AGN and SF in feedback processes. 

Optical emission lines can trace the warm-ionized phase ($T\sim10^3$--$10^4$ K) of outflows, reaching line-of-sight velocities ($LoSVs$) of $10^2$--$10^3 \ \mathrm{km \ s^{-1}}$, and spatial scales up to $\sim10^3$ pc (\citealt{2018_Cicone}, and references therein). The collisionally-excited [\ion{O}{3}]$\lambda 5007$ emission line---weakly affected by blending with nearby lines, and usually presenting high signal-to-noise ratio ($S/N$) in AGN---is a popular tracer of outflows, used to study possible connections with winds in other spectral bands  \citep[e.g.,][]{2013_Mullaney, 2017_Perna, 2018_Venturi}

\citet{2016_Woo} studied outflow kinematics with [\ion{O}{3}]$\lambda 5007$ and H$\alpha$ for a large sample of type 2 AGN at $z \leq 0.3$, finding that  higher outflow velocities correspond to higher velocity dispersions and luminosities. The gas velocity and velocity dispersion were more extreme for [\ion{O}{3}]$\lambda 5007$  than for H$\alpha$, suggesting that H$\alpha$ traces the nebular emission from SF regions---with their motion dominated by the host galaxy gravitational potential---and that [\ion{O}{3}]$\lambda 5007$ traces mainly the AGN-driven outflow.

Consistent results were reported by \citet{2016_Karouzos} studying the spatially resolved kinematics of outflows in six type 2 AGN ($z \sim0.05$--$0.1$), using integral field spectroscopy (IFS) with GMOS/Gemini. They confirmed that H$\alpha$ follows the kinematics of stellar absorption lines, while [\ion{O}{3}]$\lambda5007$ has independent and more extreme kinematics. High spatial and spectral resolution IFS has boosted knowledge of geometry and physics of galaxy-scale AGN-driven outflows in nearby galaxies \citep[i.e.,][]{2017a_Lopez-Coba, 2019_Mingozzi, 2020_Lopez-Coba}.

\subsection{NGC 7469: starburst and AGN}

 NGC 7469 is a nearby galaxy hosting a Seyfert type 1 AGN with a supermassive black hole mass $\log_{10} M_{BH} = 7.32^{+0.09}_{-0.10}  \ \mathrm{M_\odot}$ and bolometric luminosity $L_{bol} = \sim 10^{45} \ \mathrm{erg \ s^{-1}}$ \citep{2012_Ponti}. It is classified as a luminous infrared galaxy (LIRG) due to the starburst concentrated in its circumnuclear ring, triggered by interaction with the IC 5283 galaxy. The ring outer radius is  $2.5''$ ($\sim 830$ pc) and the inner radius $0.7"$ ($\sim 232$ pc), with bright knots at $\sim1.5''$ ($\sim 500$ pc) \citep{1995_Genzel}. This ring contains young ($1$--$20$ Myr) massive stars \citep{2007_Diaz-Santos}.
 
 \citet{2004_Davies} and \citet{2015_Izumi} found molecular gas structures in the central region using millimeter observations, including a circumnuclear disk ($\sim300$ pc) that \citet{2020_Izumi} revealed to be an X-ray dominated region produced by the AGN.
 
\cite{2007_Blustin} reported X-ray spatially-unresolved nuclear winds (warm absorbers) with LoSVs of $-580$ to $-2300 \ \mathrm{km \ s^{-1}}$, later confirmed by \cite{2017_Behar} and \cite{2018_Mehdipour} at $LoSVs$ of $-400$ to $-1800 \ \mathrm{km \ s^{-1}}$ within a distance of $2$--$80$ pc from the black hole. All these authors reported UV counterparts of the X-ray winds. 

 	\cite{2011_Muller-Sanchez} found a biconical outflow in the infrared coronal line [\ion{Si}{6}]$\lambda1.96\mathrm{\mu m}$ using VLT/SINFONI and Keck/Osiris IFS, extending up to $380 \pm 25$ pc from the AGN, with a maximum velocity of $\sim 130 \ \mathrm{km \ s^{-1}}$ at $220 \pm 25$ pc.
 	
 	  While this outflow was not reported in recent optical IFS observations with GTC/MEGARA by \cite{2020_Cazzoli}, they described a non-rotational turbulent component that might be associated with it.

We report, for the first time, optical extended ionized outflows in the circumnuclear region of NGC 7469, based on the kinematics of the  [\ion{O}{3}]$\lambda 5007$ and H$\beta$ emission lines. The comoving distance to NGC 7469 is $69.64$ Mpc\footnote{Using the Ned Wright Cosmology Calculator \citep{2006_Wright}.} ($z = 0.01632$, \citealt{1996_Keel}), adopting a standard $\Lambda$CDM cosmology ($H_0 = 70 \ \mathrm{km \ s^{-1} \ Mpc^{-1}}$, $\Omega_\Lambda = 0.7$ and $\Omega_M = 0.3$).

\section{Methodology}
\label{sec:methods}

\subsection{Observations and data reduction}
\label{sec:datared} 

\begin{figure*}[ht!]
\plotone{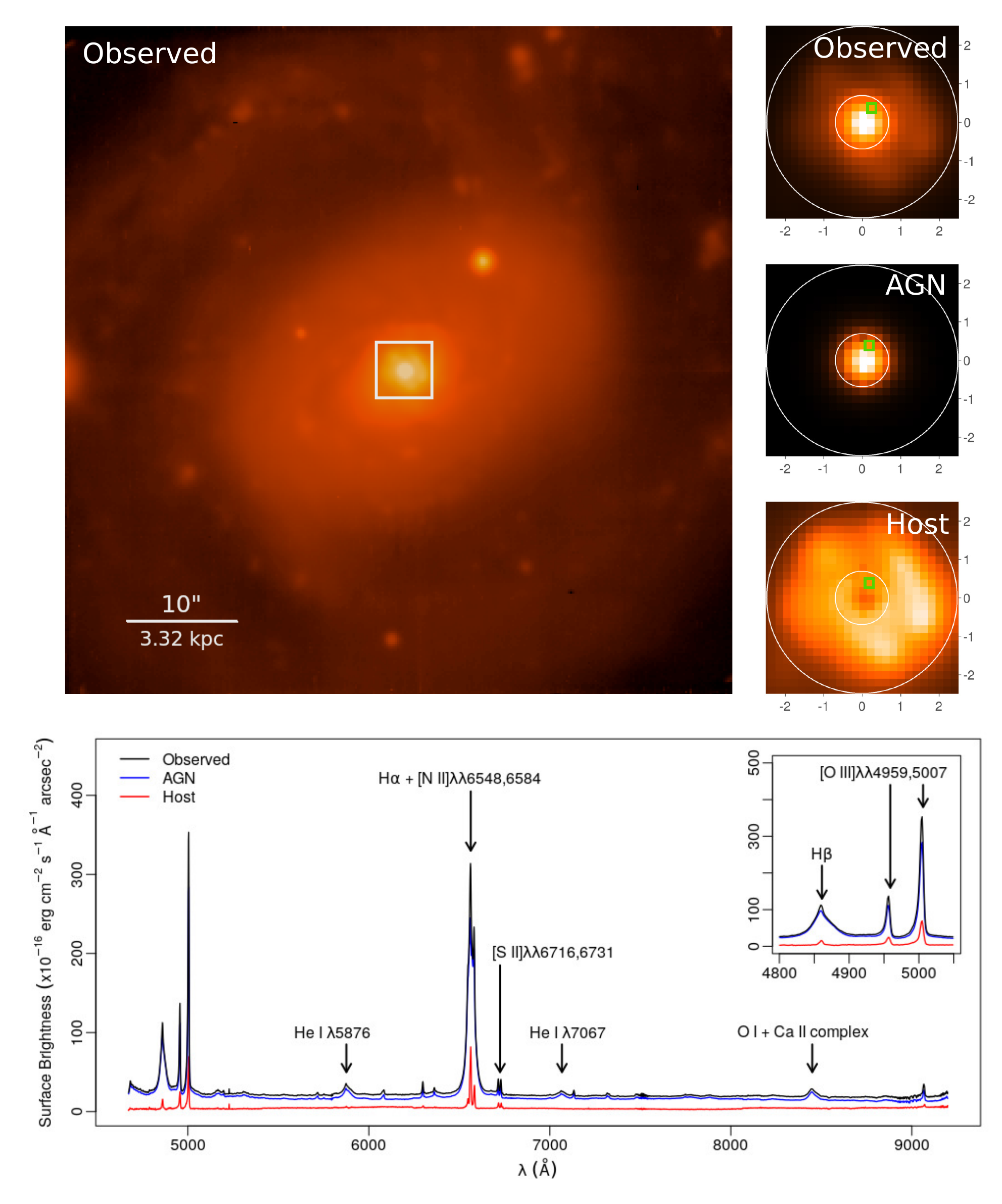}
\caption{\label{fig:Deblend}  Top-left: MUSE ``white" image (all wavelengths integrated) of NGC7469. The white square encloses the $5'' \times 5''$ field explored in this study. North is up and east is left. Right: a zoom into the $5'' \times 5''$ field for the Observed, AGN and Host data cubes resulting of the deblending process. The white circles are the inner and outer limits of the SF ring, as proposed by \cite{1995_Genzel}. Axes units are in arcseconds. Bottom: spectrum extracted from the spaxel marked in green in the right-side panels, showing the deblended spectra. Since this spaxel is close to the center ($\sim 0.4''$), the AGN dominates the flux of the spectrum.}
\end{figure*}

NGC 7469 was observed in 2014 August 19, during the science verification run of the Multi Unit Spectroscopic Explorer \citep[MUSE,][]{2004_Bacon} IFS instrument at the Very Large Telescope (VLT) of the European Southern Observatory (ESO, Chile). The pilot study of the All-weather MUse Supernova Integral-field of Nearby Galaxies survey \citep[AMUSING,][]{2015_Galbany} and the AMUSING++ compilation\footnote{\url{http://ifs.astroscu.unam.mx/AMUSING++/index.php?start=24}} \citep{2020_Lopez-Coba} include these data. MUSE covers a field of view of $1 \ \mathrm{arcmin^2}$ with a spatial sampling of  $0.2''$ per spaxel (top-left panel,  figure \ref{fig:Deblend}). For NGC 7469 each spaxel presents a scale of $66.44$ pc ($332.2 \ \mathrm{pc \ arcsec^{-1}}$). The seeing had a $FWHM = 1.23''$ ($\sim 409$ pc). Data reduction followed the standard  procedures, using the \textsc{Reflex} \citep{2013_Freudling} package and the MUSE pipeline \citep{2014_Weilbacher}. We corrected for systemic velocity  using the value by \cite{1996_Keel} ($4898 \pm 5 \ \mathrm{km \ s^{-1}}$), obtained by a combination of emission-line methods due to the absence of absorption lines in the central region.

\subsection{AGN/host galaxy deblending}
\label{sec:deblending}

Beam smearing---the scattering of light from the spatially unresolved broad-line region (BLR) and inner narrow-line region (NLR) due to seeing---can lead to overestimation of the size of the extended narrow-line region (ENLR) and of the velocity of associated  outflows in type 1 AGN \citep{2016_Husemann}. This effect was corrected through the  \textsc{QDeblend3D} software described in \cite{2012_Husemann, 2014_Husemann,2016_Husemann}. We assume a  surface brightness model for the host galaxy, fixing brightness, effective radius, Sersic index and axis ratio, with values from \cite{2009_Bentz}. \textsc{QDeblend3D} produces two data cubes (figure \ref{fig:Deblend}): one contains the AGN continuum and emission from the BLR and the NLR  (hereafter ``AGN data cube"); the other contains the host galaxy stellar continuum, SF regions and the ENLR (hereafter ``Host data cube").

\subsection{Continuum subtraction}
\label{sec:cont_subt}

 We subtract a synthetic stellar continuum template from each spectrum in the Host data cube, to obtain ``pure emission'' spectra. We create the templates (after correcting for Galactic extinction using dust maps produced by \citealt{1998_Schlegel}) using the \textsc{Starlight} population synthesis code \citep{2005_CidFernandes}. We use the base set of $150$ simple stellar populations\footnote{We use the $2016$ version of the MILES libraries, an update of ones by \citealt{2003_Bruzual}: \url{http://www.bruzual.org/bc03/Updated_version_2016/}} selected by \cite{2007_Asari}---$25$ ages ($1 \times 10^{6}$--$1.8 \times 10^{10}$ yr) and $6$ metallicities ($0.005$--$2.5 \ \mathrm{Z_\odot}$)---with the \cite{2003_Chabrier} initial mass function. Emission lines are masked out.  Intrinsic extinction is corrected in the process using the  \cite{1989_Cardelli} reddening law ($R_V = 3.1$). 

We study the line profiles of [\ion{O}{3}]$\lambda\lambda 4959,5007$ and H$\beta$ in a box  $5''$ ($1.611$ kpc) wide, centerd on the AGN, that contains the SF ring and corresponds to the field studied by \cite{2007_Diaz-Santos} (figure \ref{fig:Deblend}). The line profiles of many spaxels are asymmetric and broadened at their bases, suggesting the presence of winds. We characterize these features using two approaches (see below).

The results of [\ion{O}{3}]$\lambda 4959$ and  [\ion{O}{3}]$\lambda 5007$ are consistent, including the expected one-third flux ratio. We only report here the results for  [\ion{O}{3}]$\lambda 5007$ (hereafter [\ion{O}{3}]), due to its higher $S/N$.

\subsection{Non-parametric approach}
\label{sec:non-param-method}

\begin{figure}
\gridline{\fig{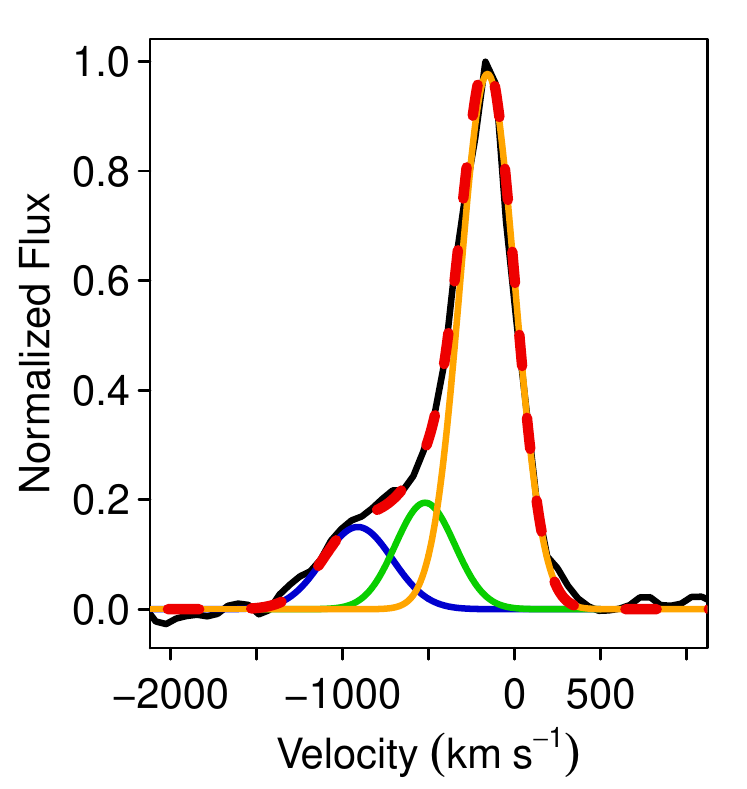}{0.35\textwidth}{(a)}}
\gridline{\fig{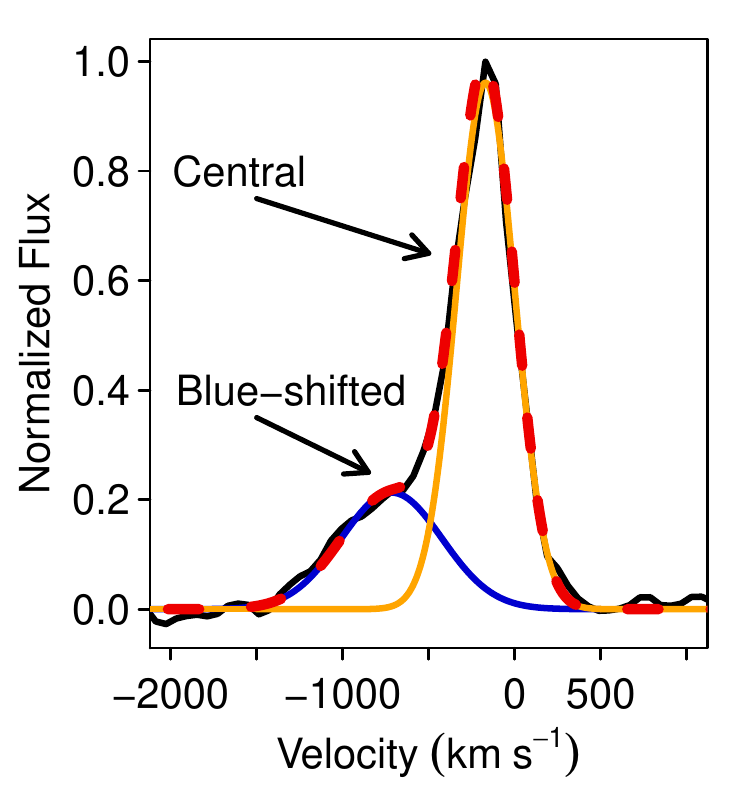}{0.35\textwidth}{(b)}}
\caption{\label{fig:Gaussians} Examples of (a) three-Gaussian fit and (b) two-Gaussian fit of [\ion{O}{3}]$\lambda 5007$. The solid black line represents the Host data cube spectrum after subtracting the stellar continuum. The red-dashed line is the synthetic line profile resulting from the sum of the components.}
\end{figure}

 Following \cite{2014_Harrison}, each emission line is fitted with three Gaussian components (figure \ref{fig:Gaussians}a), whose sum creates a synthetic line profile. To fit the Gaussians we use the  Levenberg-Marquardt algorithm \citep{1963_Marquardt}, implemented with the IDL MPFIT libraries \citep{2009_Markwardt}. Spectra were interpolated, resulting in $\approx  50$ data points per emission line, versus $\approx 10$ in the original data. This increases the chance of obtaining a good solution by decreasing the solution space, and reduces computing time. We reject all Gaussian components with $S/N < 3$.

 We avoid physically interpreting the Gaussians. Instead we characterize the  kinematics by deriving quantities from the cumulative function of the whole synthetic line profile: the offset velocity $\Delta v$ (the mean of the velocities at the 5th and 95th percentiles) the width at $80 \%$ of the flux $W_{80}$, and the $FWHM$. Hence $\Delta v$ measures the asymmetry of the line profile related with gas motion on the line of sight; the sign indicates the direction. The $W_{80}/FWHM$ ratio is the relative broadening at the base of the line.

\cite{2020_Lopez-Coba} uses an alternative non-parametric approach, however the current approach is accurate enough for our goal.

\subsection{Two-Gaussian components approach}
\label{sec:Gauss-method}

We fit two Gaussian components to the [\ion{O}{3}] line (figure \ref{fig:Gaussians}b) following the approach by \cite{2016_Woo} and \cite{2016_Karouzos}.  The Gaussian component closest to the rest-frame velocity (hereafter, \textit{central component}), is related with gas dominated by the host galaxy gravitational potential, rotating in the galactic disk; the second Gaussian component accounts for the outflowing gas as a whole (hereafter, \textit{blueshifted component}). See figure \ref{fig:Gaussians}b.

 We apply the same fitting method described in section \ref{sec:non-param-method}. We measure the kinematic parameters of each Gaussian component: the line-of-sight velocity ($LoSV$) from the Doppler shift of the Gaussian peak with respect to the rest frame, and the velocity dispersion ($\sigma$) from the $FWHM$---corrected for the instrumental width ($FWHM_{inst} \sim 158 \ \mathrm{km \ s^{-1}}$ for [\ion{O}{3}]).

\subsection{Uncertainties} 
\label{sec:uncertainties}  

We estimate uncertainties through Monte Carlo simulations, following \cite{1992_Lenz}, iterating $1000$ times. For the non-parametric approach, the mean uncertainties across the studied field are: $\sim 3.9$, $\sim 6.7$ and $\sim 6.6 \ \mathrm{km \ s^{-1}}$ for $\Delta v$, $W_{80}$ and the $FWHM$ of [\ion{O}{3}], respectively. For H$\beta$ the corresponding values are $\sim 2.9$, $\sim 6.5$ and $\sim 6.6 \ \mathrm{km \ s^{-1}}$.  

For the two-Gaussian components approach, the mean uncertainties in $LoSV$ and $\sigma$ are, respectively, $\sim 60$ and $\sim 32 \ \mathrm{km \ s^{-1}}$ for the [\ion{O}{3}] blueshifted component and $\sim 30$ and $\sim 6.6 \ \mathrm{km \ s^{-1}}$ for the  central component.

\section{Results}
\label{sec:results}

\subsection{Non-parametric approach}
\label{sec:results_nonparam}

\begin{figure*}
\gridline{\fig{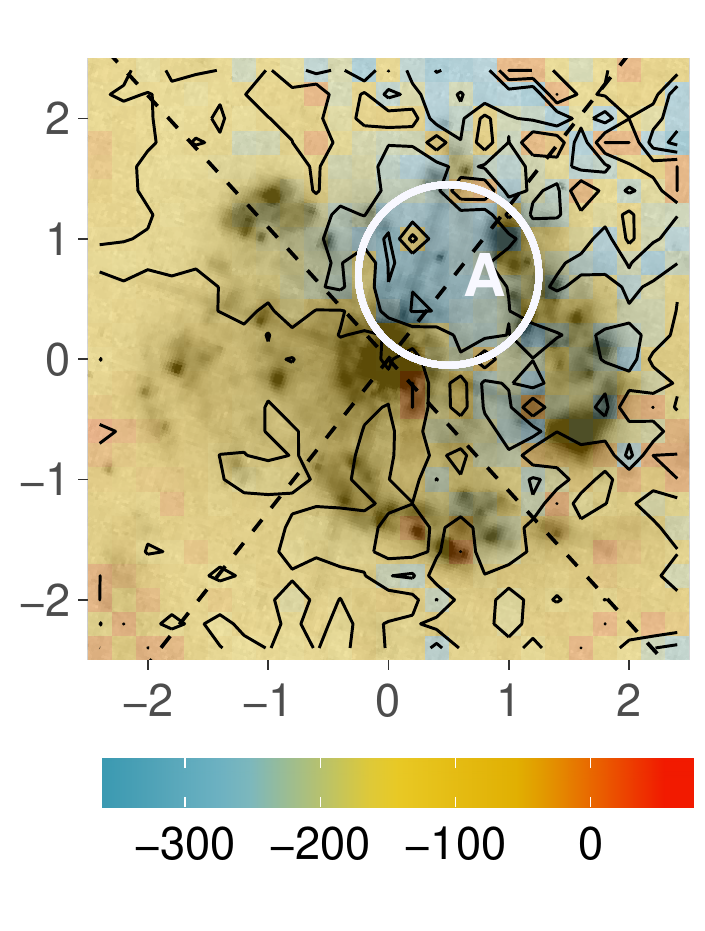}{0.3\textwidth}{(a) [\ion{O}{3}] $\Delta v$ ($\mathrm{km \ s^{-1}}$)}
          \fig{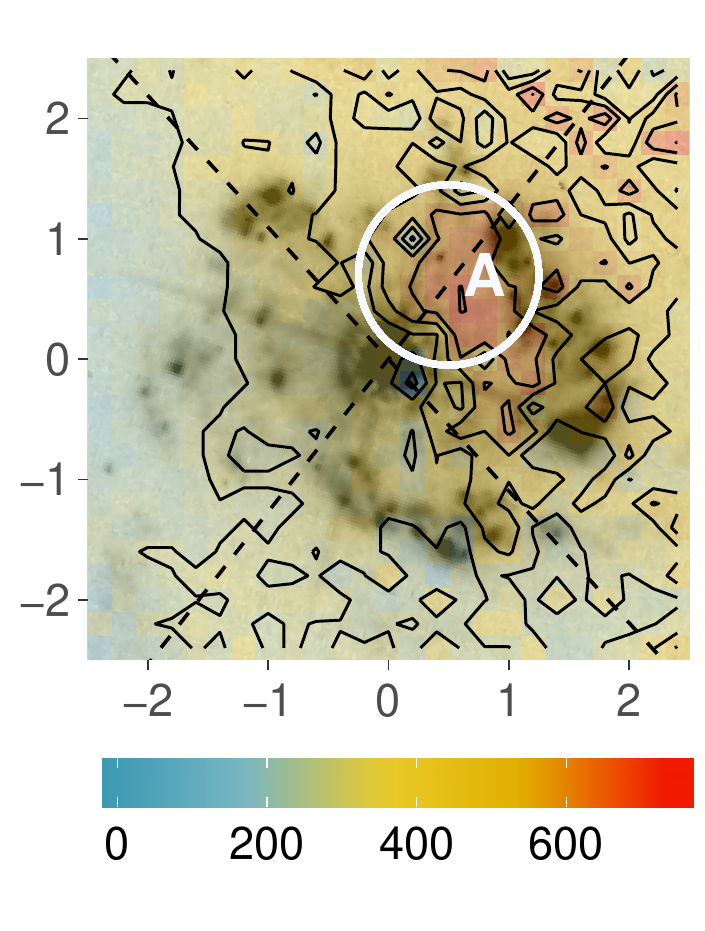}{0.3\textwidth}{(b) [\ion{O}{3}] $W_{80} \ \left(\mathrm{km \ s^{-1}}\right)$} 
          \fig{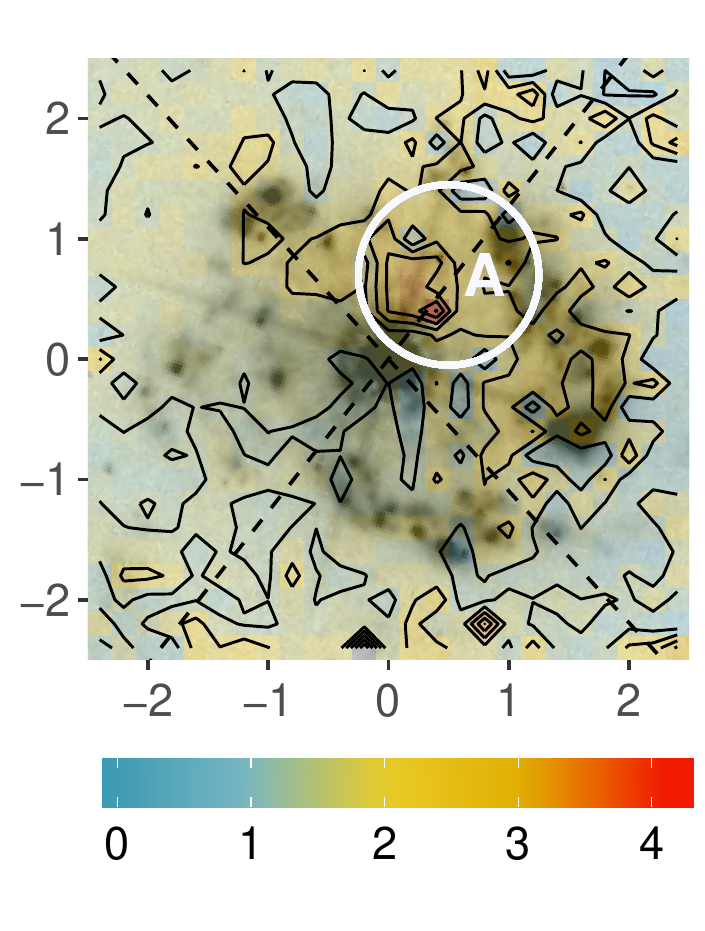}{0.3\textwidth}{(c) [\ion{O}{3}] $W_{80}/ FWHM$}
}
\gridline{\fig{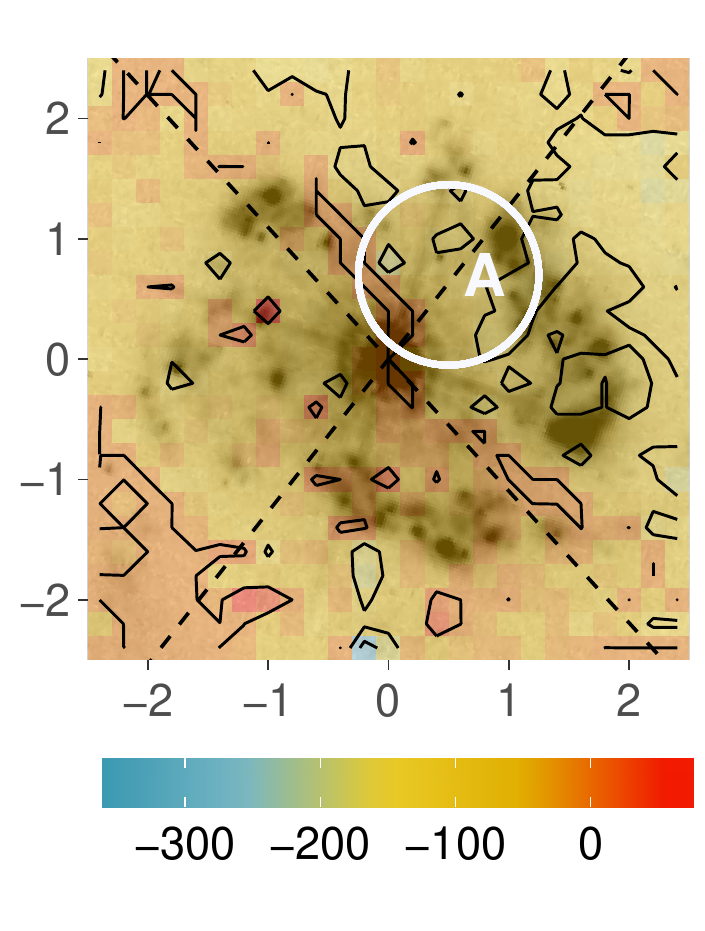}{0.3\textwidth}{(d) H$\beta$ $\Delta v$ ($\mathrm{km \ s^{-1}}$)}
          \fig{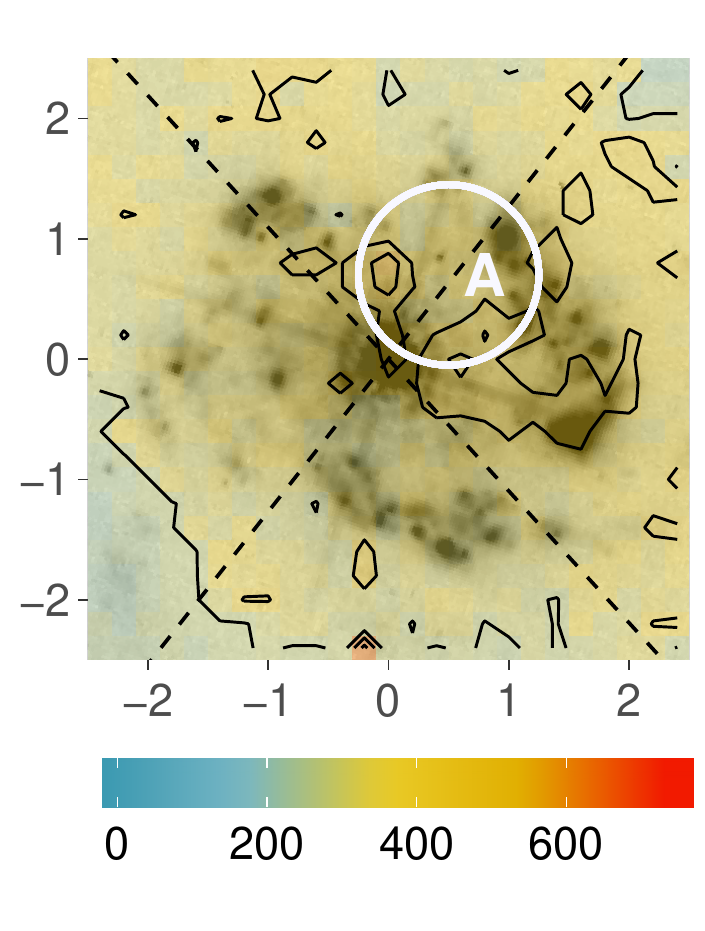}{0.3\textwidth}{(e) H$\beta$ $W_{80} \ \left(\mathrm{km \ s^{-1}}\right)$}
          \fig{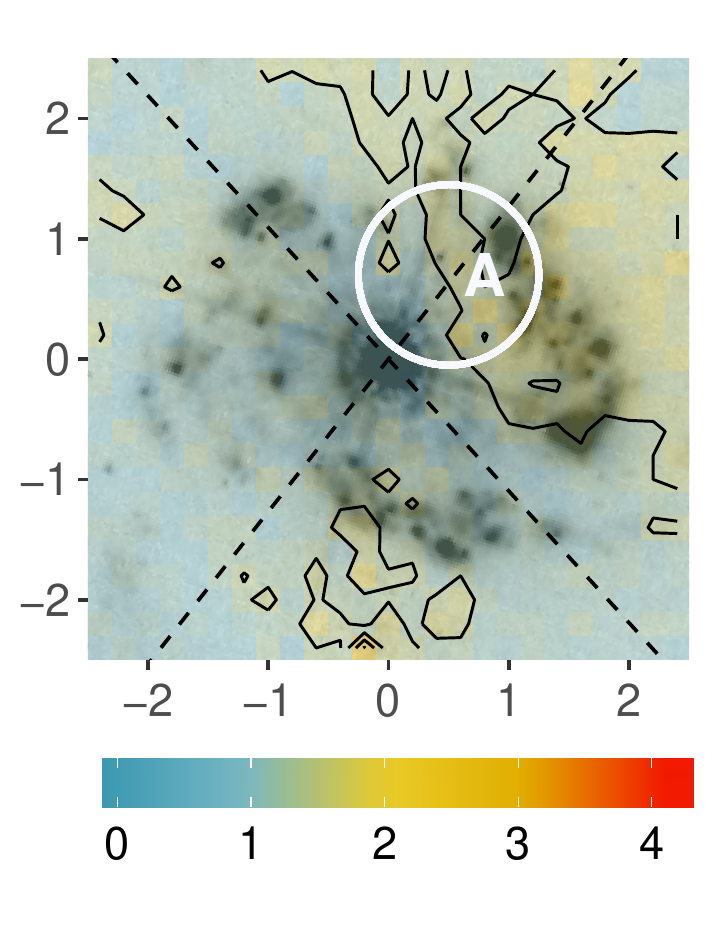}{0.3\textwidth}{(f) H$\beta$ $W_{80}/ FWHM$}
}
\caption{\label{fig:nonparam_maps} Maps from the non-parametric approach for [\ion{O}{3}] (a, b and c) and H$\beta$ (d, e and f), superimposed over the HST/ACS F330W image. North is up and east is left, axes units are in arcseconds. The AGN is located at the center. The diagonal dashed lines represent the projection on the plane of the sky of the biconical [\ion{Si}{6}]$\lambda 1.96 \mathrm{\mu m}$ outflow, with the blue side pointing west and the red side pointing east. The approximate position of the fast [\ion{O}{3}] outflow is marked with a white ``A'' and a circle (radius and position are merely orientative).}
\end{figure*}

Figure \ref{fig:nonparam_maps} shows the maps of $\Delta v$, $W_{80}$ and $W_{80}/FWHM$ for [\ion{O}{3}] and H$\beta$. As a reference for the position of the AGN and the massive SF regions of the ring, the maps are overlapped to the Hubble Space Telescope (HST) ACS F330W near-UV image\footnote{HSTScI public archive through the MAST web tool.}. Projections of the edges of the [\ion{Si}{6}]$\lambda1.96\mathrm{\mu m}$ outflow by \cite{2011_Muller-Sanchez} are shown, with the blueshifted cone pointing west and the red-shifted cone pointing east. 

 The [\ion{O}{3}] results reveal the existence of two outflow kinematic regimes, located in different regions. In figure \ref{fig:nonparam_maps}a, in a region labelled as ``A'', a strong [\ion{O}{3}] blueshifted asymmetry ($\Delta v$ up to $ -310 \ \mathrm{km \ s^{-1}}$, reaching $\sim 531$ pc ($\sim 1.6''$) to the north) extends northwest of the center, in between the AGN and the massive SF regions of the ring. Approximately half of it extends within the limits of the projected west [\ion{Si}{6}]$\lambda 1.96 \mathrm{\mu m}$ cone, while the rest lies outside to the north. 

This region also presents a broadening of $W_{80} > 600 \ \mathrm{km \ s^{-1}}$ and $W_{80}/FWHM > 3$ (figures \ref{fig:nonparam_maps}b and 3c). This high asymmetry and broadening are strong kinematic evidence of the presence of an outflow in the ``A" region. 

The existence of a second kinematic regime in the rest of the ring and the inner regions is disclosed by less prominent asymmetry ($-200 \leq \Delta v \leq -100 \ \mathrm{km \ s^{-1}}$) and broadening ($250 < W_{80} < 500 \ \mathrm{km \ s^{-1}}$ and $1.5 < W_{80}/FWHM < 3$). We ignore here the outer regions of the field due to their lower $S/N$.

The H$\beta$ maps show only one outflow regime, similar to the slowest one observed in [\ion{O}{3}]. The $\Delta v$, $W_{80}$ and $W_{80}/FWHM$ maps are shown in figures \ref{fig:nonparam_maps}d, \ref{fig:nonparam_maps}e and \ref{fig:nonparam_maps}f, respectively. Outside the ``A" region, H$\beta$ has similar but less pronounced $\Delta v$  than [\ion{O}{3}], with most spaxels having $\Delta v >-200 \ \mathrm{km \ s^{-1}}$. 

Low $W_{80}/FWHM < 2$ ratios dominate the field except for the western SF ring. Despite reaching $W_{80} > 300 \ \mathrm{km \ s^{-1}}$ (with a maximum of $\sim 640 \ \mathrm{km \ s^{-1}}$), they keep $W_{80}/FWHM < 3$. 
High $FWHM$ values keep low $W_{80}/FWHM$ ratios, possibly due  to the presence of turbulence or shocks, not only in the wind but also in the bulk of the gas, affecting the whole line profiles.

\subsection{Two-Gaussians approach}
\label{sec:Results_Gauss}

\begin{figure*}
\gridline{\fig{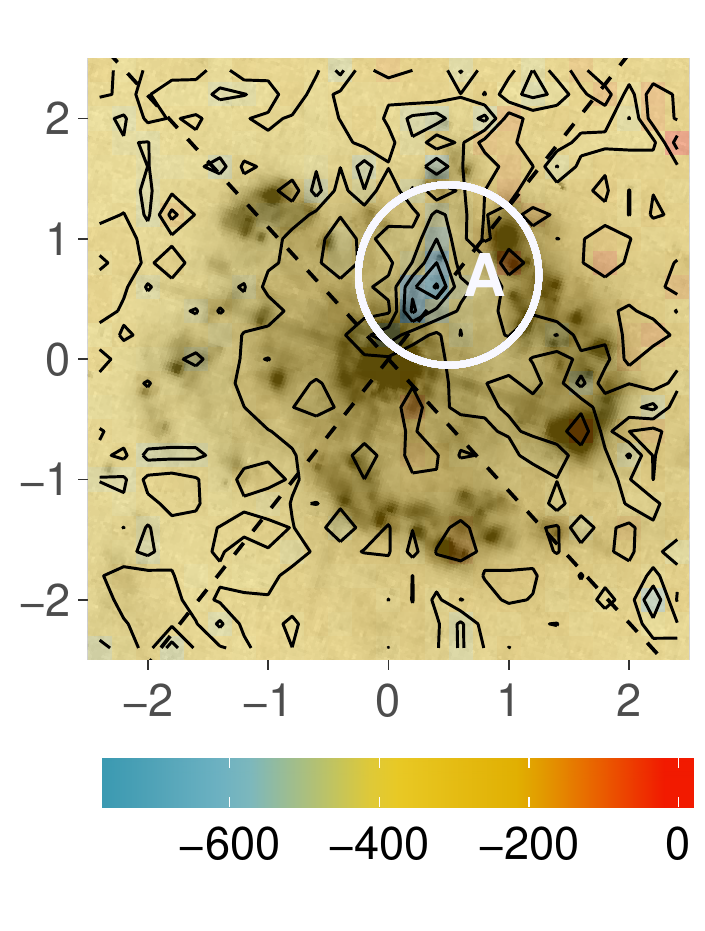}{0.3\textwidth}{(a) [\ion{O}{3}] blueshifted component $LoSV \ \left(\mathrm{km \ s^{-1}}\right)$}
          \fig{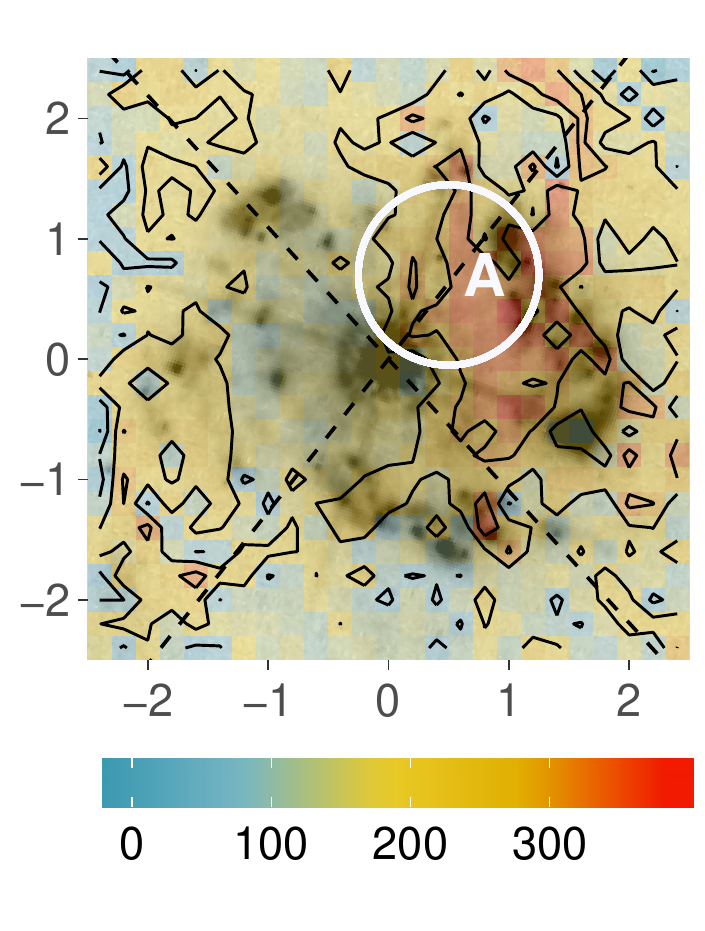}{0.3\textwidth}{(b) [\ion{O}{3}] blueshifted component $\sigma \ \left(\mathrm{km \ s^{-1}}\right)$}
          \fig{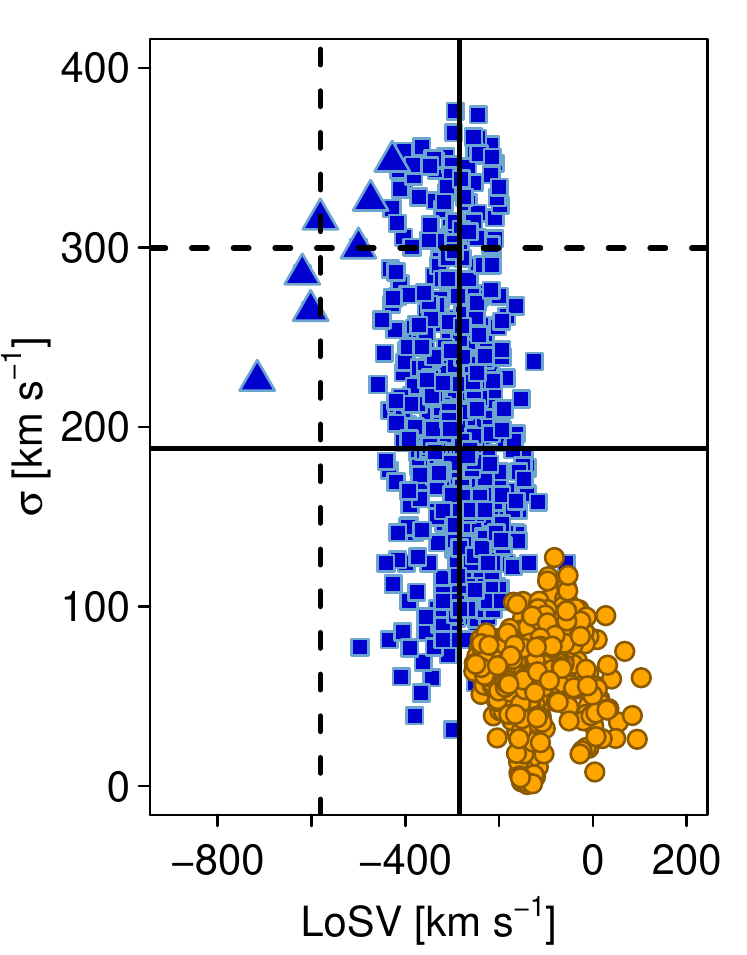}{0.3\textwidth}{(c) [\ion{O}{3}] VVD diagram}
          }                          
\caption{\label{fig:comp_maps} Maps of (a) $LoSV$ and (b) $\sigma$ for the blueshifted component of [\ion{O}{3}]. Description is the same as in figure \ref{fig:nonparam_maps}. (c) VVD diagram for [\ion{O}{3}]. The symbols correspond to the central component (orange circles), blueshifted component (blue squares) and the high-velocity blueshifted component ($LoSV < -400 \ \mathrm{km \ s^{-1}}$) spaxels in the ``A'' region (blue triangles). The dashed lines are the median values for the high-velocity spaxels while the solid lines are the median values for the rest of the blueshifted components. The central component has a mean $LoSV =  -134 \ \mathrm{km \ s^{-1}}$. Should this value be assumed as the systemic rest frame, the existence of both outflow regimes would still hold since at least two Gaussians are necessary to fit the line profiles.}
\end{figure*}

Figure \ref{fig:comp_maps}a shows the $LoSV$ map of the [\ion{O}{3}] blueshifted component. The two outflow regimes found in Section \ref{sec:results_nonparam} are confirmed, with different $LoSV$ ranges of the blueshifted component: high velocities ($LoSV < -400 \ \mathrm{km \ s^{-1}}$) are found in the ``A'' region; lower velocities ($LoSV > -400 \ \mathrm{km \ s^{-1}}$) are found around the rest of the field. The high $LoSV$ spaxels in figure \ref{fig:comp_maps}a, cover a smaller area than that of the high $\Delta v$ spaxels in figure \ref{fig:nonparam_maps}a.  The fastest blueshifted component (shown at figure \ref{fig:Gaussians}b) reaches $LoSV = -715 \ \mathrm{km \ s^{-1}}$, at $\sim 240$ pc ($\sim 0.7''$) northwest of the center. Velocity dispersion values of $\sigma$ ($ > 300 \ \mathrm{km \ s^{-1}}$) do appear in the the ``A'' region (figure \ref{fig:comp_maps}b), similar to $W_{80}$, but they extend further over the western massive SF regions.

The velocity--velocity dispersion (VVD) diagram \citep{2016_Woo, 2016_Karouzos} provides a straightforward visualization of the different kinematic regimes (figure \ref{fig:comp_maps}c).  Once the central and blueshifted components are plotted in the VVD, the former is clearly is separated from the latter in the parameter space.  Following \cite{2016_Woo} and \cite{2016_Karouzos} interpretation, the more extreme kinematics of the blueshifted component could be evidence of outflows (not dominated by the gravitational potential of the host galaxy), while the positive and negative $LoSV$ values of the central component suggest motion in a galactic disk.

The high $LoSV$ spaxels in the ``A'' region of figure \ref{fig:comp_maps}a (with a median $LoSV$ of $-581 \ \mathrm{km \ s^{-1}}$ and median $\sigma$ of $300 \ \mathrm{km \ s^{-1}}$) are clearly separated from the bulk of the blueshifted components (with median $LoSV$ and $\sigma$ of $-284$ and $188 \ \mathrm{km \ s^{-1}}$, respectively) consistent with the existence of the two kinematic regimes in the blueshifted component.

Note that if the mean $LoSV$ of the central component was assumed as systemic reference frame, the outflow velocity would be $134 \ \mathrm{km \ s^{-1}}$ slower. However, the existence of both outflow regimes would still hold solidly, with the slower gas mean $LoSV \sim -150 \mathrm{km \ s^{-1}}$. In fact, more than one Gaussian is needed to fit the line profiles, as shown by the non-parametric approach.

\subsection{BPT diagnostics of the central component}

The BPT diagnostic diagram \citep{1981_Baldwin} informs on the excitation mechanisms, provided that only  Gaussian components with similar kinematics, i.e. tracing the same bulk of gas, are considered. The blueshifted component kinematics of [\ion{O}{3}] and H$\beta$ are inconsistent for many spaxels, and shall not be combined in the same diagram. Due to blending of the shifted components, we could unambiguously identify only the peaks of the central components in the H$\alpha$-[\ion{N}{2}] complex and use them for BPT diagnostics.

\begin{figure*}
\gridline{\fig{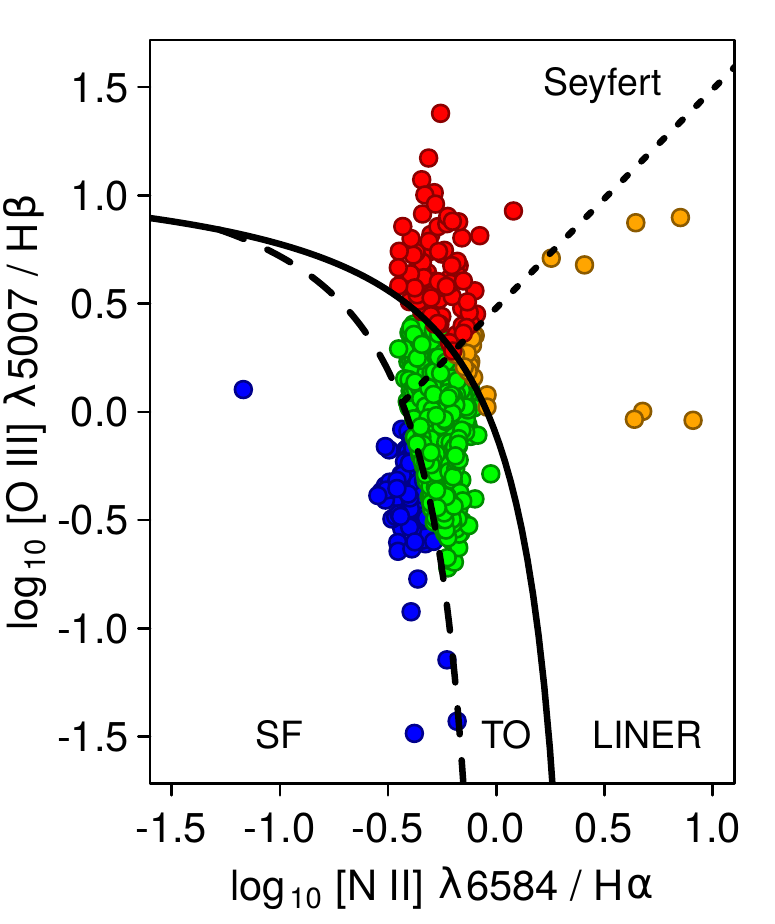}{0.4\textwidth}{(a)}
          \fig{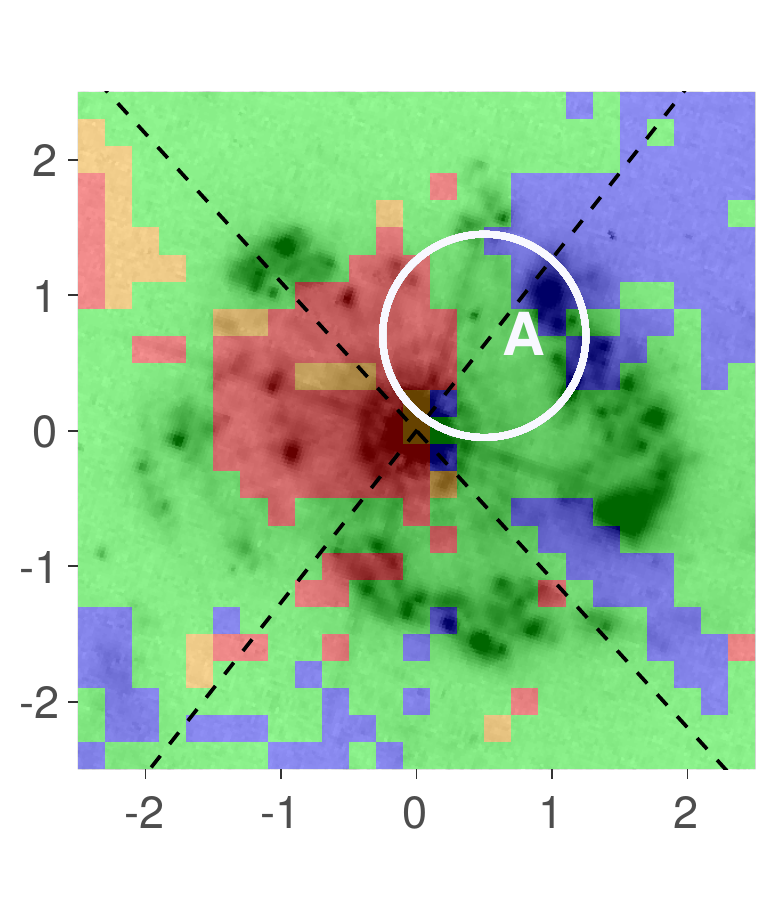}{0.4\textwidth}{(b)}
          }                          
\caption{\label{fig:BPT} (a) BPT-NII diagram of the central component, used to determine the main excitation mechanism in each spaxel: star formation (SF), transition object (TO), low ionization nuclear emission line region (LINER) and Seyfert. The dividing lines correspond to \cite{2001_Kewley} (solid), \cite{2003_Kauffmann} (dashed) and \cite{2010_CidFernandes} (dotted). (b) BPT-NII map, the color of each spaxel corresponds to its position in the diagnostic diagram. Other elements are the same as in figure \ref{fig:nonparam_maps}.}
\end{figure*}

Figure \ref{fig:BPT} shows the BPT-NII diagnostics---based on the [\ion{O}{3}]$\lambda 5007 / \mathrm{H\beta}$ and [\ion{N}{2}]$\lambda 6584 / \mathrm{H\alpha}$ line ratios---of the central component. The mean flux uncertainties (in $\times 10^{-17} \ \mathrm{erg \ cm^{-2} \ s^{-1}}$ units) are $\sim 0.58$ for [\ion{O}{3}], $\sim 0.52$ for H$\beta$, $\sim 0.76$ for [\ion{N}{2}]$\lambda 6584$ and $\sim 3.4$ for H$\alpha$.

AGN excitation dominates the northeast quadrant up to $\sim 500$ pc from the center, while a combination of SF and transition-object-like (TO) excitation extends across the rest of the field (LINER-like excitation is scarce). The ``A" region  with the fastest [\ion{O}{3}] outflow  presents a combination of AGN, TO and SF excitation, the latter over the massive SF regions of the ring. We use this result as exploratory; direct extension to the blueshifted component should be avoided. More detailed analysis will be presented in a future paper.

\section{discussion}
\label{sec:discussion}

Both kinematic analyses indicate two [\ion{O}{3}] outflow regimes: the ``A'' region shows more extreme [\ion{O}{3}] outflow kinematics compared to the rest of the field (figures \ref{fig:nonparam_maps}a,  \ref{fig:nonparam_maps}b, \ref{fig:nonparam_maps}c, \ref{fig:comp_maps}a, \ref{fig:comp_maps}b); this is reflected in the VVD diagram (figure \ref{fig:comp_maps}c).  

H$\beta$ and [\ion{O}{3}] winds behave similarly across the field, except for the ``A'' region, where the behaviour of [\ion{O}{3}] is more extreme. The H$\beta$ maps (figures \ref{fig:nonparam_maps}d, \ref{fig:nonparam_maps}e and \ref{fig:nonparam_maps}f) show a slight increment in $\Delta v$ and $W_{80}/FWHM$ in the ``A'' region, but not as pronounced as with [\ion{O}{3}]. 

 Based purely on kinematic criteria, a stellar origin for the slow H$\beta$ and [\ion{O}{3}] outflows would be consistent with them extending across most of the SF ring.  The AGN is probably driving the faster [\ion{O}{3}] outflow regime: it presents more extreme kinematics, with similar $LoSVs$ to those AGN-driven outflows reported by \cite{2016_Woo} and \cite{2016_Karouzos} in [\ion{O}{3}] (projection effects on the $LoSV$ should be considered).

 The high  $\sigma$ values of the [\ion{O}{3}] blueshifted component and the low H$\beta$ $W_{80}/FWHM$ ratios (but high $W_{80}$ values) at the ``A'' region suggest the presence of shocks both in the wind and in the gravitationally-bounded gas. Therefore, interaction between the AGN-driven and SF-driven winds in this region is a possibility.
 
  As for the excitation mechanisms, the BPT-NII map of the central component (gas in the galactic disk) in figure \ref{fig:BPT}b shows that emission in most of  region ``A" and the north-east quadrant is consistent with a combination of AGN and SF excitation, while the rest of the field  is consistent with SF excitation (although AGN contribution cannot be excluded).
  
  The fast [\ion{O}{3}] outflow partially overlaps with the blueshift cone of the AGN-driven outflow traced by [\ion{Si}{6}]$\lambda 1.96 \mathrm{\mu m}$. If both oxygen and silicon were photoionized by the AGN, the geometry of the AGN radiation field could determine the spatial distribution of their emission, with [\ion{Si}{6}]$\lambda 1.96 \mathrm{\mu m}$ being detected where the density of high-energy photons was higher---Si$^{+5}$ has a higher ionization potential than O$^{++}$ ($167$ versus $35.1$ eV).
  
  All this makes plausible the following scenario: the slow outflow could be driven by SF regions while the fast outflow could be driven by the AGN.   However, evidence is not conclusive. BPT diagnostics of the blueshifted component will test our proposed scenario by providing insights on the excitation mechanisms in the outflowing gas. We are working on that analysis; results will be presented in a future paper.
 
The beam-smearing correction here applied for the first time to NGC 7469 \citep[e.g.][]{2020_Cazzoli, 2020_Lopez-Coba}, proved to be crucial to remove the AGN spectral contribution, allowing detection of the outflows.

Confirmation that these outflows were driven by the SF and the AGN, would make of NGC 7469---a galaxy close enough to spatially distinguish their sources---an outstanding case for studying combined feedback effects.

\section*{Acknowledgements}

We thank the referee for the careful review of our paper and insightful comments that have improved the quality of our work. Support from CONACyT (Mexico) grant CB-2016-01-286316 is acknowledged. We are thankful to Bernd Husemann for helping with \textsc{QDeblend3D} installation. ACRO thanks  Vital Fern\'andez for advice and discussions. JPTP acknowledges DAIP-UGto (Mexico) for granted support (0173/2019). YA aknowledges support from project PID2019-107408GB-C42 (Ministerio de Ciencia e Innovaci\'on, Spain). SFS thanks the support of CONACYT grants CB-285080 and FC-2016-01-1916, and funding from the PAPIIT-DGAPA-IN100519 (UNAM) project. LG was funded by the European Union's Horizon 2020 research and innovation programme under the Marie Sk\l{}odowska-Curie grant agreement No.839090. This work has been partially supported by the Spanish grant PGC2018-095317-B-C21 within the European Funds for Regional Development (FEDER). 

{\bf \software{REFLEX \citep{2013_Freudling}, MUSE pipeline \citep{2014_Weilbacher}, Starlight \citep{2005_CidFernandes}, QDeblend3D \citep{2012_Husemann}, MPFIT \citep{2009_Markwardt}}}



\bibliographystyle{aasjournal}
\bibliography{Bibliography}





\end{document}